\documentclass[aps,twocolumn,prd,superscriptaddress,nofootinbib,floatfix]{revtex4-2}

\usepackage{mathtools}
\usepackage{amsfonts}
\usepackage{amssymb}
\usepackage{mathrsfs}
\usepackage{bbm}
\usepackage{slashed}
\usepackage{amsmath}
\usepackage{bm}
\usepackage{bbold}

\usepackage{graphicx}
\usepackage{color}
\usepackage{colortbl}
\usepackage{array}

\usepackage{float}
\usepackage{placeins}
\usepackage{booktabs}
\usepackage[caption=false]{subfig}
\usepackage{makecell}
\usepackage{tabstackengine}

\usepackage{xspace}
\usepackage{hyperref}
\usepackage[nameinlink]{cleveref}
\usepackage{bookmark}
\usepackage{siunitx}

\usepackage{xifthen}
\usepackage{xcolor}
\hypersetup{
	colorlinks,
	linkcolor={red!75!black},
	citecolor={blue!75!black},
	urlcolor={blue!75!black}
}

\usepackage[utf8]{inputenc}
\allowdisplaybreaks[4]

\newcommand{\feyn}[1]{
	\setbox0=\hbox{\ensuremath{#1}}
	\hbox to\wd0{\hbox to0pt{\hbox to\wd0{\hss/\hss}\hss}\box0}}

\def\Eq#1{\Cref{#1}}
\def\Fig#1{\Cref{#1}} 
\def\Tab#1{\Cref{#1}}
\setkeys{Gin}{width=0.48\textwidth}
\graphicspath{{./figures/}}

\newcommand{\gettitle}{}

\newcommand{\getGiessenAffiliation}{\affiliation{Institut f\"ur Theoretische Physik, Justus-Liebig-Universit\"at Gie\ss en, 35392 Gie\ss en, Germany}}

\hypersetup{
	colorlinks,
	linkcolor={red!75!black},
	citecolor={blue!75!black},
	urlcolor={blue!75!black},
	pdftitle={\gettitle},
	pdfauthor={Zelle},
	pdfkeywords={effective theory} {analytic continuation}
	{correlations functions} {hadronization}
	{functional renormalisation group}
	{real time} {bound states}
	bookmarksopen=true,
	bookmarksopenlevel=2,
	bookmarksnumbered=true
}

\begin{document}

\title{Mapping the critical region along the second-order chiral phase boundary}

\author{Shi Yin}
\email{Shi.Yin@theo.physik.uni-giessen.de}
\getGiessenAffiliation

\begin{abstract} 

We investigate the extent of the critical scaling region of the chiral phase transition at finite chemical potential within the quark–meson (QM) model using the functional renormalization group (fRG) approach. By analyzing the scaling behavior of the chiral order parameter and correlation length with respect to temperature and pion mass near the second-order phase transition, we extract critical exponents from the data and quantify the range over which the scaling relations remain valid. We find that both the leading order and the next-to-leading-order scaling regions systematically shrink as the chemical potential increases. This behavior is observed in both the local potential approximation (LPA) and its extension including anomalous dimensions ($\mathrm{LPA}^\prime$), with qualitatively consistent results, while the scaling region in $\mathrm{LPA}^\prime$ is slightly smaller than that in LPA.

\end{abstract}
	
\maketitle

\section{Introduction}
According to the quantum chromodynamics (QCD), strong interaction matter undergoes a transition from a hadronic phase to a quark–gluon plasma (QGP) phase at finite temperature. For physical values of the current quark masses, this transition is a smooth crossover \cite{HotQCD:2018pds,Borsanyi:2020fev}. In the limit of massless light quarks, i.e., chiral limit, the crossover is replaced by a genuine chiral phase transition. As predicted in \cite{Pisarski:1983ms}, in the chiral limit of light quarks, the transition may be associated with a second-order phase transition in the $3d-O(4)$ universality class. However, due to the impact of the $U_A(1)$ anomaly on the phase transition, determining the Columbia plot of the QCD phase diagram remains an active topic of research. To further investigate the nature of the QCD phase transition, numerous model studies \cite{Chen:2021iuo,Schaefer:2004en,Pawlowski:2014zaa,Giacosa:2024orp,Fejos:2024bgl}, functional QCD studies \cite{Braun:2020ada,Gao:2021vsf,Braun:2023qak} and lattice QCD simulations \cite{HotQCD:2019xnw} have been carried out.

At present, there is no consensus on the size of the critical region associated with the chiral phase transition. In recent years, both lattice QCD and functional continuum approaches have provided estimates of the size of this critical region. The behavior of different thermodynamic observables is analyzed by lattice QCD, such as the pseudo-critical temperature at different pion masses in an attempt to extract critical scaling behavior \cite{HotQCD:2019xnw,Borsanyi:2020fev,Aarts:2020vyb,Kotov:2021rah,Aarts:2023vsf,Ding:2023oxy}. Although functional continuum approaches are limited by truncations, their relatively precise calculations near the chiral limit provide valuable insights into the study of the chiral phase transition. The magnetic equation of state for 2+1-flavor QCD has been computed by Dyson–Schwinger equation (DSE) \cite{Gao:2021vsf}. The results provided a first estimate of the size of the critical region along the mass direction and also determined the order of the phase transition on the Columbia plot \cite{Bernhardt:2025fvk}. Subsequently, the functional renormalization group (fRG) has been applied to QCD to perform high-precision calculations of the critical behavior in the chiral limit \cite{Braun:2020ada,Braun:2023qak}. Near the chiral critical temperature of QCD in the chiral limit, fRG yields critical exponents associated with the Wilson–Fisher (WF) fixed point of the $O(4)$ universality class. They further investigate the pion mass range over which the magnetic equation of state obeys critical part of scaling function, finding that the scaling behavior is only valid for pion mass below approximately 10 MeV. This mass range is outside the data range of lattice QCD. Both DSE and fRG studies suggest that, at vanishing chemical potential, the critical region along the pion mass direction is smaller than anticipated and dominated by soft modes.

To circumvent the complexity of full QCD computation while preserving its essential features, several low-energy effective models, such as $O(N)$ model \cite{Juttner:2017cpr,Tan:2022ksv}, linear sigma model \cite{Umekawa:1999qx,Grahl:2013pba}, quark–meson (QM) model \cite{Stokic:2010piu,Berges:1997eu,Braun:2010vd,Chen:2021iuo}, and the Nambu–Jona-Lasinio (NJL) model \cite{Strouthos:2005ks,Fu:2007xc}, have been employed to study the chiral phase transition. Non-equilibrium dynamical critical behavior has likewise been studied using the fRG approach, see \cite{Chen:2024lzz,Roth:2023wbp}. While critical exponents have been widely studied in models and QCD, existing analyses are largely restricted to vanishing chemical potential, and the dependence of the critical region on chemical potential remains poorly explored. 

In the chiral limit, the chiral phase transition at finite chemical potential remains of second-order until it turns into a first-order transition at the tri-critical point (TCP). The TCP connects the $O(4)$ second-order transition to the first-order one and is also linked to the $Z(2)$ second-order transition line, along which the critical endpoint (CEP) at physical quark mass lies, see \cite{Chen:2021iuo}. By estimating how the size of the critical region varies with chemical potential, one can obtain a rough expectation for the extent of the critical region around the TCP and, by extension, the CEP. One can thereby infer the possible range over which critical fluctuations associated with a CEP in the QCD phase diagram may affect experimental observables, e.g., baryon number fluctuations \cite{Fu:2023lcm,Fu:2021oaw}. For studies of the critical behavior of the QM model at the TCP and CEP, analyzed via quark-number cumulants within mean-field and fRG approaches, see \cite{Schaefer:2011ex,Schaefer:2006ds}. 

In this work, we perform a first exploratory study using a QM model computed within the fRG framework. Following a strategy similar to that of \cite{Braun:2023qak}, we compute the chiral order parameter and the correlation length, and examine the ranges in temperature and pion mass where their behaviors are governed by the critical part of the scaling function, then investigate the density dependence of the size of the static critical region.

This paper is organized as follows: In \Cref{sec:QM_frg}, we briefly introduce the QM model employed in this work and the flow equations involved in the fRG approach. In \Cref{sec:exp}, the order parameter and correlation length are computed around the phase transition, and their temperature and pion mass dependence are investigated along the phase boundary. The range of applicability of the critical part of the scaling functions is discussed in \Cref{sec:phaseboundary}. In the end, the summary and conclusions are given in \Cref{sec:conclusion}.
\section{quark-meson model within functional renormalization group approach}
\label{sec:QM_frg} 
As a starting point, this section introduces the QM model employed in this work, along with the technical details required for the fRG approach calculations. In order to incorporate the QM model into the framework of fRG, we first formulate the scale-dependent effective action of the model
%
\begin{align}\label{eq:action}
\Gamma_k[\phi,q,\bar{q}]&=\int_x \bigg\{
 Z_{q,k}\,\bar q\big[\gamma_\mu \partial_\mu-\gamma_0\,\hat \mu\big]q + \frac{1}{2}\,Z_{\phi,k}\,(\partial_\mu\phi)^2\nonumber\\[2ex]
&+\frac{1}{2}\,h_k\,\bar q(\tau^0 \sigma+i\gamma_5 \boldsymbol{\tau}\cdot\boldsymbol{\pi})q+U_k(\rho)-c\sigma\bigg\}\,.
\end{align}
%
In this effective action, we have the degrees of freedom of the hadrons $\phi=(\sigma,\boldsymbol{\pi})$ and the (anti-) quark ($\bar q$) $q$. The range of the Lorentz index $\mu$ is taken to be from 0 to 3. For simplicity, we consider two light-quark flavors with degenerate masses and chemical potentials, i.e., $m_q\equiv m_u=m_d$ and $\hat\mu=\mathrm{diag}(\mu,\mu)$. The interaction strength between mesons and quarks is given by the Yukawa coupling $h_k$. Then we have the tensor $\tau^0=\mathbbm{1}_{N_f\times N_f}$ and Pauli matrices $\boldsymbol{\tau}$ with $N_f=2$ and $N_c=3$. The effective potential $U_k(\rho)$ captures the chiral phase transition at finite temperature and density together with the linear chiral symmetry breaking term $-c\sigma$. The $c$ in the linear term gives the strength of the chiral symmetry breaking which is related to the current light quark mass. When we set this strength to zero, the system reaches the chiral limit. Note that in this work our computations are carried out within both the Local Potential Approximation (LPA) and $\mathrm{LPA}^\prime$. In LPA, the scale dependence of the meson (quark) wave function renormalization $Z_{\phi/q,k}$ and the Yukawa coupling $h_k$ will not be taken into account. In $\mathrm{LPA}^\prime$, we take into account the flow of meson wave function renormalization $Z_{\phi,k}$, i.e., meson anomalous dimension. In both cases we set the $Z_{q,k}=1$, because the quark wave function renormalization does not alter the critical exponents of the $O(4)$ universality class.

To solve the effective action at infrared energy scale, we have to introduce the Wetterich equation \cite{Wetterich:1992yh}
%
\begin{align}\label{eq:wetterich_eq}
\partial_t\Gamma_k[\phi,q,\bar q]=\frac{1}{2}\mathrm{Tr}\bigg(G^{\phi\phi}_{k}\partial_t R^\phi_{k}\bigg)-\mathrm{Tr}\bigg(G^{q\bar q}_{k}\partial_t R^q_{k}\bigg)\,.
\end{align}
%
In order to preserve the dimensional consistency of the equations, we use the renormalization group (RG) time $t=\mathrm{ln(k/\Lambda)}$. Here $\Lambda$ gives us the ultraviolet (UV) scale of our current theory, we set it to 500 MeV in this work. The index $\phi$ and $q$ stand for the boson and quark d.o.f. in our system. The trace is performed on all the intrinsic degrees of freedom, e.g., flavor, color and Lorentz space. The $G^{\phi\phi/q\bar{q}}_k$ is the propagator and the $R^{\phi/q}_k$ is the regulator of meson/quark, the definitions of them are given in \Cref{app:flow_eq}.

We start with the flow equation of the effective potential
%
\begin{align}\label{eq:u_flow}
\partial_t U_k(\rho)=&\frac{k^4}{4\pi^2}\bigg[(N_f^2-1)l^{(\phi,4)}_0(m^2_{\pi,k},\eta_{\phi,k};T)\nonumber\\[2ex]
&+l^{(\phi,4)}_0(m^2_{\sigma,k},\eta_{\phi,k};T)\nonumber\\[2ex]
&-4N_cN_fl^{(q,4)}_0(m^2_{q,k};T,\mu)\bigg]
\end{align}
%
The $l^{(\phi/q,4)}_0$ is the threshold function of a bosonic field or a quark field, the expression is also given in \Cref{app:flow_eq}. In addition, they are all functions of meson masses or quark masses. Then the dimensionless masses are defined by the derivative of the effective potential with respect to the field
%
\begin{align}
\bar m^2_{\pi,k}&=\frac{U_k'(\rho)}{k^2Z_{\phi,k}}\,,\\[2ex]
\bar m^2_{\sigma,k}&=\frac{U_k'(\rho)+2\,\rho\,U_k''(\rho)}{k^2Z_{\phi,k}}\,,\\[2ex]
\bar m^2_{q,k}&=\frac{h_k^2\rho}{2\,k^2\,Z_{q,k}}\,,
\end{align}
%
with the $O(4)$ invariant meson field $\rho=\phi^2/2$. The meson anomalous dimension is given by $\eta_{\phi,k}=-\partial_tZ_{\phi,k}/Z_{\phi,k}$. We use the Taylor expansion method to solve the effective potential. We give the expansion form of the potential
%
\begin{align}\label{eq:taylor}
U_k(\rho)=\sum^N_{n=0}\frac{\bar\lambda_{n,k}}{n!}(\bar\rho-\bar\kappa_k)^n\,.
\end{align}
%
The expansion coefficients, meson field and the expansion point are all RG-invariant quantities. They are defined by $\bar\lambda_{n,k}=\lambda_{n,k}/Z^n_{\phi,k}$, $\bar{\rho}=\rho\,Z_{\phi,k}$ and $\bar{\kappa}_k=\kappa_k\,Z_{\phi,k}$. In this work, we expand the potential up to $N=5$, in \cite{Yin:2019ebz,Pawlowski:2014zaa} we can see that under this order sufficient convergence can already be observed. The flow equations of the Taylor coefficients can be obtained by the field derivative of \Eq{eq:u_flow}
%
\begin{align}\label{eq:flow_coef}
&\partial_{\bar\rho}\Big(\partial_tU_k(\rho)\Big)\bigg|_{\rho=\kappa_k}\nonumber\\[2ex]
=&(\partial_t-n\,\eta_{\phi,k})\bar{\lambda}_{n,k}-(\partial_t\bar{\kappa}_k+\eta_{\phi,k}\bar{\kappa}_k)\bar{\lambda}_{n+1,k}\,.
\end{align}
%
The flow of the expansion point $\partial_t\bar{\kappa}_k$ is given by the gap equation of the potential
%
\begin{align}\label{eq:gapeq}
\partial_{\bar\rho}\Big(U_k(\rho)-\bar c_k\sqrt{2\bar\rho}\Big)\bigg|_{\rho=\kappa_k}=0\,.
\end{align}
%
The scale dependence of the explicit chiral symmetry breaking term is only given by the meson wave function renormalization $\bar c_k=c/\sqrt{Z_{\phi,k}}$, and $\partial_t\bar c_k=\eta_{\phi,k}\bar c_k/2$. We solve the \Eq{eq:flow_coef} and \Eq{eq:gapeq} together and get the flow equation of the expansion point
%
\begin{align}
\partial_t\bar \kappa_k=&-\frac{\bar c_k^2}{\bar\lambda^3_{1,k}+\bar c_k^2\,\bar\lambda_{2,k}}\nonumber\\
&\times\bigg\{\partial_{\bar\rho}\Big(\partial_tU_k(\rho)\Big)\bigg|_{\bar\rho=\bar\kappa_k}+\eta_{\phi,k}\Big(\frac{\bar\lambda_{1,k}}{2}+\bar\kappa_k\,\bar\lambda_{2,k}\Big)\bigg\}\,.
\end{align}
%
These equations are under the $\mathrm{LPA}^\prime$ truncation. If we set $\eta_{\phi,k}=0$ and $Z_{\phi,k}=1$ at every RG-scale, we can go back to LPA truncation.

The flow equation of meson anomalous dimension is given by a momentum derivative of the meson two-point function
%
\begin{align}\label{eq:eta_phi}
\eta_{\phi,k}=-\frac{\delta_{ij}}{3Z_{\phi,k}}\frac{\partial}{\partial(|\mathbf{p}|^2)}\frac{\delta^2\partial_t\Gamma_k}{\delta\pi_i(-p)\delta\pi_j(p)}\bigg|_{p=0}\,.
\end{align}
%
Note that, we use the spatial component of the pion wave function renormalization in our computation. This approximation is checked in \cite{Yin:2019ebz}, at low density area this setup can ensure the $O(4)$ critical behavior very well. The numerical setup of the QM model is also given in \Cref{app:numerical}.

For the Yukawa coupling $\bar h_k$, we do not consider the flow contribution of the quark-meson vertex. So the flow equations of it only comes from the meson anomalous dimension $\partial_t \bar h_k=\eta_{\phi,k}\bar h_k/2$.
\section{Critical exponents at finite chemical potential}
\label{sec:exp}
The critical exponents characterize how different thermodynamic quantities diverge near the phase transition temperature. One can obtain the exponents by different methods. The most rigorous way to compute critical exponents is to solve the RG fixed-point equations, see e.g., \cite{Tetradis:1993ts,Tan:2022ksv,Fejos:2022mso}. One can also compute them with the data obtained from the effective potential at critical temperature, see e.g., \cite{Chen:2021iuo,Braun:2023qak,Schaefer:2006ds}. In this work, we take the second method to reveal the behavior of the exponent at finite external parameters, e.g., temperature, chemical potential and current quark mass.

In this section, we compute the effective potential at finite temperature and density, then use the data to compute the critical behavior of the model at finite chemical potential. We mainly compute three critical exponents, i.e., $\delta$, $\beta$ and $\nu$. Moreover, we use these exponents to estimate the size of the static $O(4)$ critical region along the second-order phase boundary.
%
\subsection{Critical behavior}
\label{subsec:CE}
We start from the grand potential of the QM model. The potential is given by the effective potential and the chiral symmetry breaking term
%
\begin{align}
\Omega(T,\mu,c)=U_{k_\mathrm{IR}}(\bar\rho)-c\sigma\,.
\end{align}
%
The symmetry breaking strength $c$ is proportional to the current quark mass. So if we set $c$ to zero, the whole system is in the chiral limit. When we are in the critical region of the $O(4)$ second-order phase transition, the potential can be considered as the critical part of the scaling function
%
\begin{align}
f_s(t,h)\sim \Omega(\pm\epsilon_t,\epsilon_c)\,.
\end{align}
%
The $\epsilon_t$ and $\epsilon_c$ denote infinitesimal deviations of the thermodynamic potential from the chiral limit phase transition point along the reduced temperature direction and the chiral symmetry breaking direction, respectively. If the variations in the direction of temperature and the explicit symmetry breaking term are infinitesimal, the effective potential remains within the critical region. Therefore, it can still be described by the critical equation. Here we use the reduced temperature $t=(T-T_c)/T_c$ and the external field $h$ which is related to the chiral breaking strength $c$. The order parameter of the system can be computed by the scaling function with a derivative of the external field
%
\begin{align}
M=-\frac{\partial f_s}{\partial\,h}\equiv \sigma_0\,,
\end{align}
%
which is the expectation value of the light quark condensate. The critical exponents $\beta$ and $\delta$ can be computed through the order parameter
%
\begin{align}
\sigma_0(t,h=0)&\propto (-t)^\beta\,,\label{eq:beta}\\[2ex]
\sigma_0(t=0,h)&\propto h^{\frac{1}{\delta}}\label{eq:delta}\,.
\end{align}
%
These two exponents describe the behavior of the order parameter near the critical temperature and current quark mass. In addition, there is another important critical exponent $\nu$, which characterizes the divergence of the correlation length near the critical temperature
%
\begin{align}\label{eq:nu}
\xi\propto |t|^{-\nu}\,.
\end{align}
%
The correlation length $\xi$ is closely related to the mass of the $\sigma$ meson, see \Eq{eq:xi}. These three exponents are computed in this work to study the critical behavior along the second-order phase boundary at finite chemical potential. From the relations of exponents, we can directly obtain other exponents
%
\begin{align}
\gamma&=\beta\,(\delta-1)\,,\\[2ex]
\gamma&=\nu\,(2-\eta)\,,\\[2ex]
\beta&=\frac{\nu}{2}\,(d-2+\eta)\,,\\[2ex]
\delta&=\frac{d+2-\eta}{d-2+\eta}\,,\\[2ex]
\nu&=\frac{\beta}{d}\,(1+\delta)\,.
\end{align}
%
The anomalous dimension $\eta$ is also computed in the $\mathrm{LPA}^\prime$ case of this work.
\subsection{Critical temperature}
\label{subsec:CT}
%
\begin{figure}[t]
\includegraphics[width=0.48\textwidth]{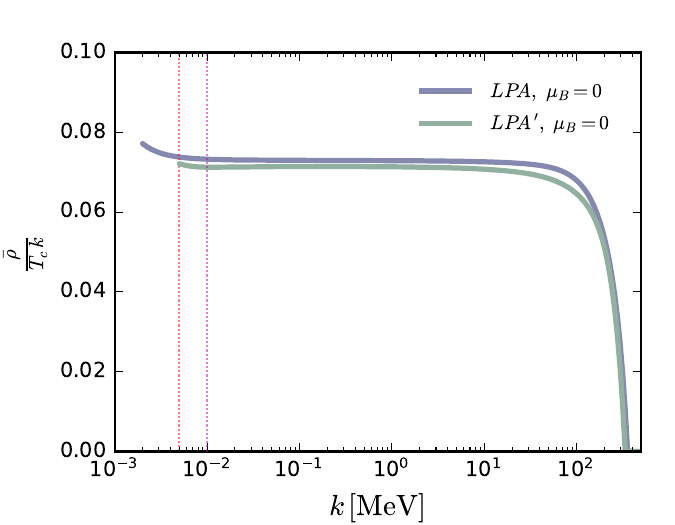}
\caption{Dimensionless $O(4)$ invariant meson field as function of RG-scale k. The red and magenta dashed lines indicate the scale k at which the LPA and $\mathrm{LPA}^\prime$ results deviate from the plateau.}
\label{fig:kappa_tc}
\end{figure}
%
We use the scaling behavior of the order parameter to locate the critical temperature along the second order phase boundary. When solving the dimensionless fRG fixed-point equations, it is common to assume that the equations are evaluated at a fixed dimensionless order parameter and to search for the corresponding fixed point of the model. Analogously, we use the dimensionless order parameter to determine whether the system is at its phase transition temperature $T_c$, corresponding to the Wilson–Fisher (WF) fixed point of a second-order transition. 

At finite temperature, we define a dimensionless effective potential by $u_k(\bar{\rho})=k^{-d}\,T^{-1}\,U_k(\bar{\rho})$. Same dimensionless method can also be found in \cite{Chen:2023tqc}. Here we have $d=3$. Then we can easily obtain the dimensionless order parameter $\bar{\rho}/(k\,T)$ by the Taylor expansion \Eq{eq:taylor}. So at the critical temperature, $\bar{\rho}/(k\,T_c)$ should be a constant and this can therefore serve as a criterion for identifying whether the model is at the critical temperature. An example is shown in \Fig{fig:kappa_tc}, where the dimensionless order parameters are plotted as functions of the RG-scale k. We observe that the order parameter vanishes in the high-scale regime and, after crossing the chiral symmetry breaking scale $k_{\chi SB}$, rapidly rises to a constant value that remains stable down to the infrared. This behavior appear in both LPA and $\mathrm{LPA}^\prime$ results. In the deep infrared region, the order parameter gradually deviates from the plateau. This deviation arises from numerical limitations, as the determination of the critical temperature becomes inaccurate at very small values of k. At this level of precision, the system gradually departs from the WF fixed point. Therefore, when computing the critical exponents, we choose the value of k within the plateau region as the infrared scale of the flow equations, as indicated by the red and magenta dashed lines in \Fig{fig:kappa_tc}.

In the following, we apply this method on different chemical potential, and the corresponding critical temperature is then determined by fine-tuning.
%
\subsection{Mass direction}
\label{subsec:delta}
%
\begin{figure*}[t]
\includegraphics[width=0.48\textwidth]{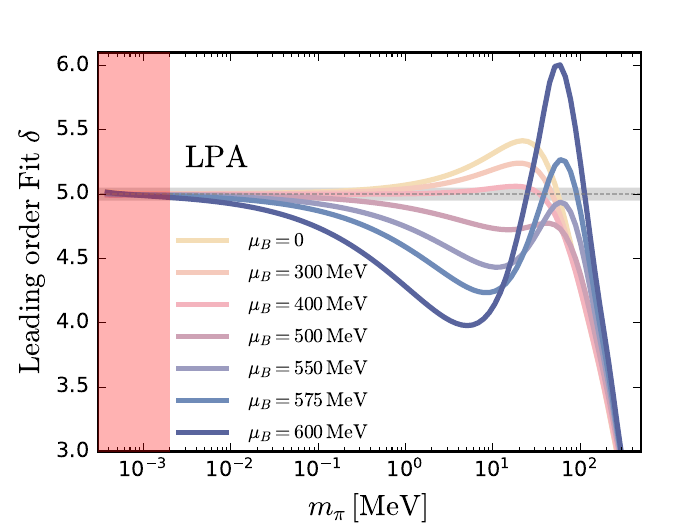}
\includegraphics[width=0.48\textwidth]{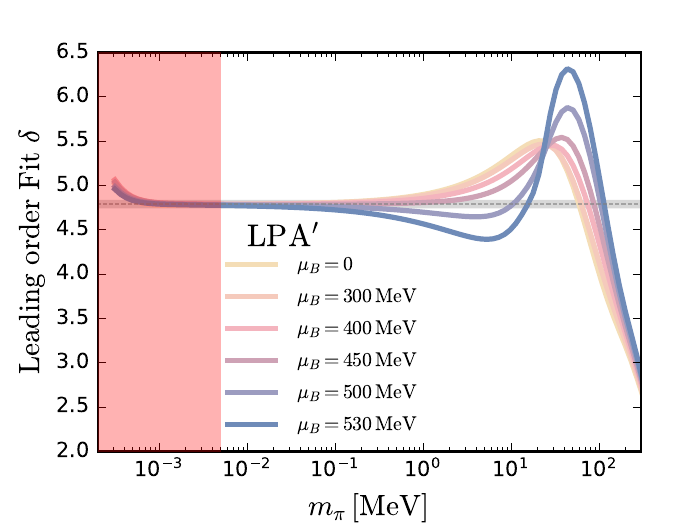}
\caption{The leading order (LO) fit of the order parameters within LPA and $\mathrm{LPA}^\prime$ under different values of pion mass. The results are obtained at $\mu_B=0,\,300,\,400,\,500,\,550,\,575,\,600$ MeV for LPA and $\mu_B=0,\,300,\,400,\,450,\,\,500,\,530$ MeV for $\mathrm{LPA}^\prime$. The right boundary of the red band is the $k_{\mathrm{IR}}$ we choose in the computation. The gray dashed line is the value of the value of the exponent, $\delta\simeq5.0$ for LPA and $\delta\simeq4.79$ for $\mathrm{LPA}^\prime$. The gray band highlights the parameter region where the deviation of the exponent reaches $1\%$.}
\label{fig:h_scaling}
\end{figure*}
%
Before we start to compute the order parameter, it must be emphasized that when solving the effective potential using the Taylor expansion method, the strict chiral limit is hard to reach due to the limitations of this approach. One may alternatively perform the Taylor expansion around different expansion points associated with the chiral restoration scale and the chiral symmetry–breaking scale \cite{Papp:1999he}. However, the determination of the matching scale $k_\chi$ between these two expansion schemes introduces additional numerical uncertainties. This renders the numerical error of this approach equivalent to that obtained in a direct calculation with a sufficiently small but non-vanishing value of c. In this work, we tune the explicit symmetry breaking term to a sufficiently small value such that it does not influence the scaling behavior of the system. The minimal pion mass accessible in our calculation is $2.7\times10^{-4}$ MeV, which is small enough to see the scaling behavior.

We first compute the order parameter at critical temperature, i.e., $t=0$, at finite chiral symmetry breaking. We use the method introduced in \Cref{subsec:CT} to determine the critical temperatures at finite chemical potential. In addition, the chiral symmetry breaking strength $c$ is proportional to the pion mass $m^2_\pi$. If we only consider the scaling part of the potential, \Eq{eq:delta} can be written as
%
\begin{align}\label{eq:mpi_log}
\mathrm{Log}\Big(\sigma_0(t=0,m^2_\pi)\Big)=\mathrm{Log}(B_c)+\delta^{-1}\,\mathrm{Log}(m^2_\pi)+\cdots\,.
\end{align}
%
We can see from the equation above, in the critical region, the critical exponent $1/\delta$ can be extracted by fitting the slope of $\mathrm{Log}(\sigma_0)$ versus $\mathrm{Log}(m^2_\pi)$. $\mathrm{Log}(B_c)$ denotes the amplitude factor of the data. The same method has also been employed in fRG calculations of QCD \cite{Braun:2023qak}. As the pion mass increases, the system gradually moves away from the $O(4)$ second-order critical region, and the slope correspondingly deviates from the value $1/\delta$.

In \Fig{fig:h_scaling}, we demonstrate the inverse of the slope of the data as functions of pion mass at different baryon chemical potentials for both LPA and $\mathrm{LPA}^\prime$. The red band gives us the unreliable region of the data. In this region the pion mass is smaller than the infrared cutoff $k_{\mathrm{IR}}$ used in the calculation, causing the data to fall outside the range of numerical accuracy. We can see that, in the small pion mass region, the fitted slope exhibits a plateau at the value corresponding to the critical exponent $\delta$. We get the critical exponent $\delta_{\mathrm{LPA}}\simeq5.00$ and $\delta_{\mathrm{LPA}'}\simeq4.79$. For zero chemical potential, the results start to deviate from the $1\%$ vicinity of the critical exponent at the pion mass below 10 MeV, which is consistent with the QCD fits including Next-leading-order (NLO) contributions \cite{Braun:2023qak}, same results are also given in \Cref{sec:subleading}. It then exhibits a peak and rapidly decreases as the mass increases. The values of the exponents obtained in this work coincide with the results of $O(N)$ model within FRG approach \cite{Litim:2001hk,Bohr:2000gp}.

At high chemical potential, the data gradually change from an upward deviation to a downward deviation. In other words, the range over which the slope agrees with the critical exponent first increases and then decreases. At very high chemical potential, the LPA fit quickly deviates from the critical exponent, while the $\mathrm{LPA}^\prime$ fit deviates slightly more slowly, however, the overall behavior remains similar to that of LPA. The maximal chemical potentials considered in the two approximations are not the same. This is because the phase structure in the chiral limit differs between the two approximations. In the $\mathrm{LPA}^\prime$, the tri-critical point is located at a smaller chemical potential, and therefore the calculations within this approximation are terminated at a lower chemical potential.

It should be emphasized that results obtained in the region where the pion mass is smaller than the infrared RG-scale cannot be considered reliable. This is because, in this region, the parameters lie beyond the range of control of the fRG calculation, and it can no longer be guaranteed that the system remains at the WF fixed point. Therefore, in the figures we highlight this region in red to guide the reader.
%
\subsection{Lower temperature direction}
\label{subsec:beta}
%
\begin{figure*}[t]
\includegraphics[width=0.48\textwidth]{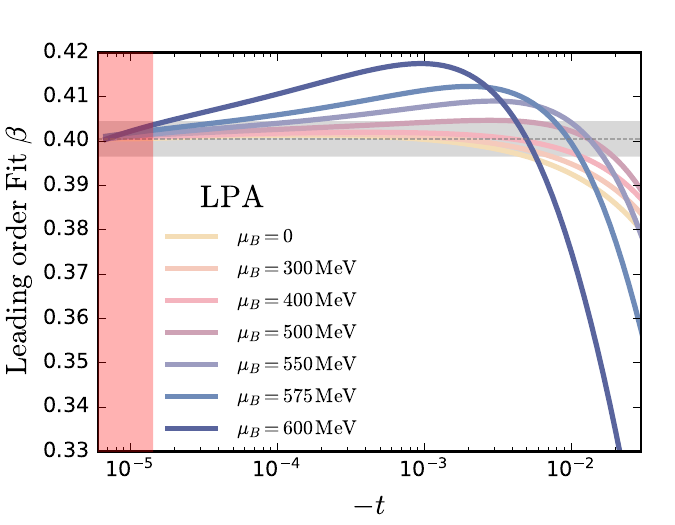}
\includegraphics[width=0.48\textwidth]{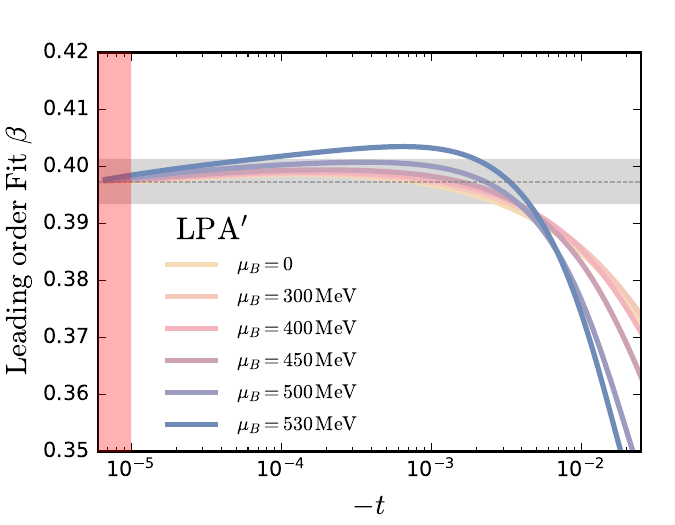}
\caption{The leading order (LO) fit of the order parameters within LPA and $\mathrm{LPA}^\prime$ under different values of temperature. The results are obtained at $\mu_B=0,\,300,\,400,\,500,\,550,\,575,\,600$ MeV for LPA and $\mu_B=0,\,300,\,400,\,450,\,500,\,530$ MeV for $\mathrm{LPA}^\prime$. The gray dashed line is the value of the exponent, $\beta\simeq0.400$ for LPA and $\beta\simeq0.397$ for $\mathrm{LPA}^\prime$. The gray band highlights the parameter region where the deviation of the exponent reaches $1\%$.}
\label{fig:t_scaling}
\end{figure*}
%
Now we change from the mass direction to the temperature direction. Along this direction, one can start from phase transition line either to the high-temperature chiral symmetry restored phase or to the low-temperature chiral symmetry broken phase. In this subsection, we first investigate the behavior of the order parameter in the low-temperature region, i.e., the critical exponent $\beta$.

To get the order parameter as a function of temperature, we first fix the model at the  (approximate) chiral limit with $m_\pi\simeq2.7\times10^{-4}$ MeV. At this mass, the scaling behavior of the order parameter shown in \Fig{fig:kappa_tc} can already be observed. Subsequently, we gradually lower the temperature starting from the critical temperature, i.e., the reduced temperature t is decreased from zero. From \Eq{eq:beta}, we can obtain the relation
%
\begin{align}\label{eq:t_log}
\mathrm{Log}\Big(\sigma_0(t,h=0)\Big)=\mathrm{Log}(A)+\beta\,\mathrm{Log}(-t)+\cdots\,.
\end{align}
%
The slope of the $\mathrm{Log}(\sigma_0)$ as function of $\mathrm{Log}(-t)$ is the exponent $\beta$. $\mathrm{Log}(A)$ is the amplitude factor along the temperature direction.

In \Fig{fig:t_scaling}, we show the results of the slope fitted by \Eq{eq:t_log}. The red band is given by the RG-time $k_{\mathrm{IR}}/\Lambda$, which shows the reliable boundary of the reduced temperature. We get the exponent $\beta_{\mathrm{LPA}}\simeq0.400$ and $\beta_{\mathrm{LPA}'}\simeq0.397$. On the temperature direction, the slope of the data is more flatten than the mass direction. For both LPA and $\mathrm{LPA}^\prime$ results, the slope stays in the gray band in a wide range. In the lower-temperature region, i.e. for larger $|t|$, the $\mathrm{LPA}^\prime$ results deviate from the gray band more rapidly than those of LPA. This can be attributed to the temperature dependence introduced by the anomalous dimension. As the chemical potential increases, the range over which the slope lies within the gray band first expands and then shrinks. For the results at high chemical potential, we find that the slope of the data rapidly deviates from the critical exponent within a very narrow range. This indicates that, at high chemical potential, the model quickly departs from the leading-order (LO) scaling function assumed in \Eq{eq:t_log}. To properly describe the model in this regime, additional regular corrections must be included, see, e.g., \cite{Braun:2023qak}.
%
\subsection{Higher temperature direction}
\label{subsec:nu}
%
\begin{figure*}[t]
\includegraphics[width=0.48\textwidth]{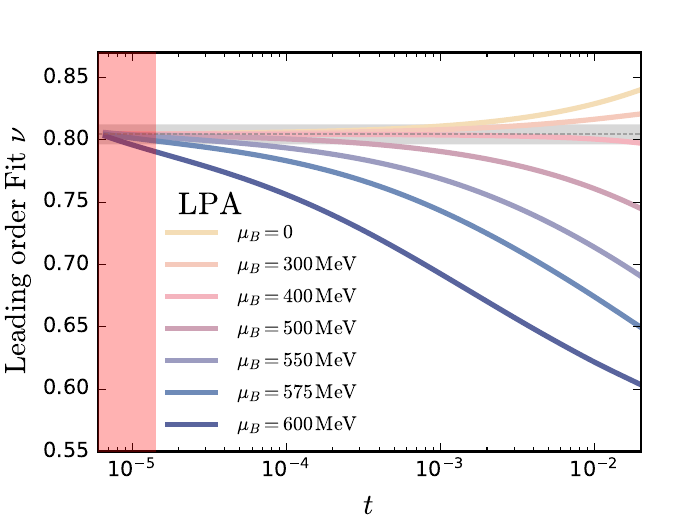}
\includegraphics[width=0.48\textwidth]{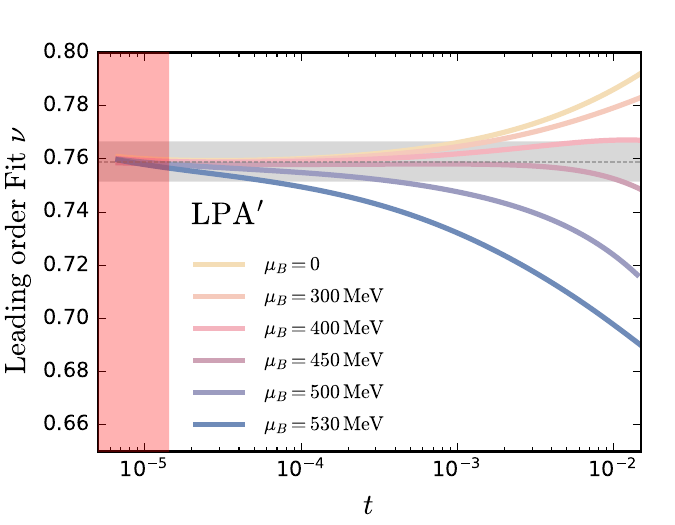}
\caption{The leading order (LO) fit of the order parameters within LPA and $\mathrm{LPA}^\prime$ under different values of temperature. The results are obtained at $\mu_B=0,\,300,\,400,\,500,\,550,\,575,\,600$ MeV for LPA and $\mu_B=0,\,300,\,400,\,450,\,500,\,530$ MeV for $\mathrm{LPA}^\prime$. The gray dashed line is the value of the exponent, $\nu\simeq0.804$ for LPA and $\nu\simeq0.759$ for $\mathrm{LPA}^\prime$. The gray band highlights the parameter region where the deviation of the exponent reaches $1\%$.}
\label{fig:xi_scaling}
\end{figure*}
%
After the lower temperature direction, we go to the other side of the phase boundary. Since in the chiral symmetry phase the order parameter is always zero, we turn to compute the critical exponent $\nu$, which is related to the correlation length of the system. From the scaling analysis, the correlation length in the $O(N)$ type model is related to the inverse mass of the $\sigma$ mode, i.e.,
%
\begin{align}\label{eq:xi}
\xi\propto\frac{1}{m_\sigma}\,.
\end{align}
%
Because in the high temperature region the chiral symmetry is restored, the sigma mass and pion mass are the same. So here we also have $\xi\propto1/m_\pi$. So from \Eq{eq:nu} we can have the scaling relation of the data
%
\begin{align}\label{eq:xi_log}
\mathrm{Log}\Big(\xi(t,h=0)\Big)=\mathrm{Log}(C)-\nu\,\mathrm{Log}(t)+\cdots\,.
\end{align}
%
The factor $\mathrm{Log}(C)$ is the amplitude of the data and $\nu$ is the critical exponent. 

In \Fig{fig:xi_scaling}, we show the slope of $\mathrm{Log}(m_\sigma)$ as function of $\mathrm{Log}(t)$ at finite reduced temperature. At low chemical potential, the behavior of the slope is similar to that of the order parameter slope in \Fig{fig:t_scaling}: it remains within a $\pm1\%$ window around the critical exponent $\nu_{\mathrm{LPA}}\simeq0.804$ and $\nu_{\mathrm{LPA}'}\simeq0.759$ over a wide range when moving away from the phase transition temperature. At finite chemical potential, the slope gradually changes from an upward deviation to a downward deviation from the critical exponent. At an intermediate value of the chemical potential, the entire curve lies within the critical exponent error window. For LPA it is around $\mu_B\sim400$ MeV, and for $\mathrm{LPA}^\prime$ is around $\mu_B\sim450$ MeV. It can be anticipated that, if one attempts to determine the size of the critical region by the deviation of the fitted slope from the critical exponent, the resulting critical region would become very large in this case.
\section{Deviation from the second-order phase boundary}
\label{sec:phaseboundary}
In this section, we summarize the numerical results obtained in the previous section and examine the rate at which the fitted slopes of the data deviate from the critical exponents along the second-order phase boundary.

In \Fig{fig:h_scaling,fig:t_scaling,fig:xi_scaling}, we fit the data obtained from QM model with LPA and $\mathrm{LPA}^\prime$ using the scaling form \Eq{eq:mpi_log,eq:t_log,eq:xi_log}. Our goal is to examine the temperature or mass ranges in which the slopes deviate from the critical exponents, thereby inferring how the size of the second-order phase transition critical region evolves with increasing chemical potential. To more clearly illustrate the deviation of the slope at finite density, we plot contour lines of the existing slope data on the phase diagram, allowing for a clearly analysis of how the deviation range evolves along the phase boundary.

\begin{figure*}[t]
\includegraphics[width=0.48\textwidth]{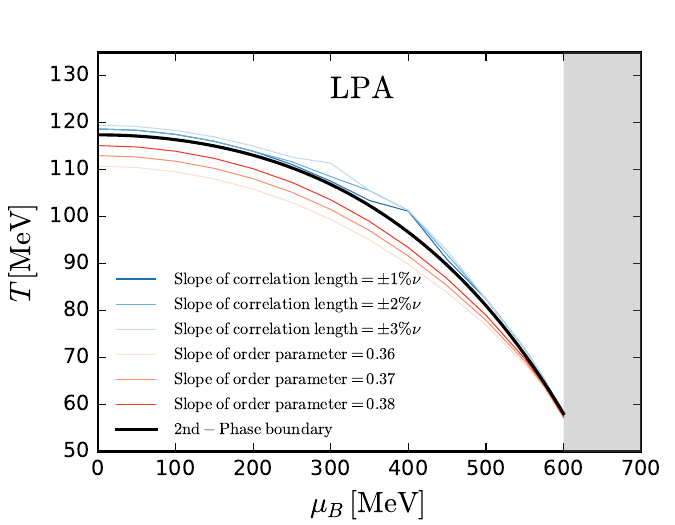}
\includegraphics[width=0.48\textwidth]{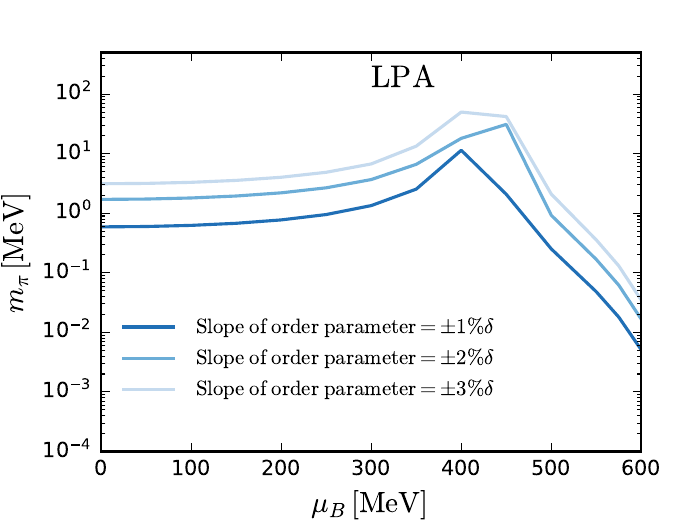}
\caption{Contour lines of the fitted LPA data slope on the phase diagram. In the left panel, the contour lines above the phase boundary (black solid line) indicate where the slope of the correlation length as a function of reduced temperature deviates from the critical exponent $\nu$ by $1\%$, $2\%$, and $3\%$. The lines below the phase boundary corresponding to the values 0.36, 0.37 and 0.38 of the slope of the order parameter $\sigma_0$ as a function of reduced temperature. In the right panel, the lines corresponding to $1\%$, $2\%$, and $3\%$ deviations of the slope of the order parameter as a function of pion mass from the critical exponent $\delta$. The gray area denotes regions where no data are available.}
\label{fig:phase1}
\end{figure*}

In \Fig{fig:phase1}, we show the results of LPA. The left plot is the phase diagram of the model which shows the contour lines along the temperature direction. For the region above the critical temperature, we extract the slope by fitting the correlation-length data and display contour lines corresponding to deviations of $1\%$, $2\%$, and $3\%$ from the critical exponent $\nu\simeq0.8$. From the figure, we observe that at low chemical potential and in the vicinity of the TCP, the slope deviates from $\nu$ very rapidly, as indicated by the dense contour lines near the phase boundary in these regions. In contrast, around a chemical potential of about 400 MeV, the slope remains close to $\nu$ over a much broader range, and the corresponding contour lines merge at a relatively large distance above the phase boundary. This region also corresponds to the transition regime in the left plot of \Fig{fig:xi_scaling}, where the slope changes from an upward deviation to a downward deviation.

In the low-temperature region, the contour lines are determined by the slope of the order parameter as a function of the reduced temperature. To improve readability and avoid the curves lying too close to the phase boundary, we choose contour lines at the three values 0.36, 0.37, and 0.38. The evolution of these three contour lines is more monotonic: as the chemical potential increases, they gradually approach the phase boundary. In the vicinity of the TCP, they almost coincide with the phase boundary, indicating that at high chemical potential the slope deviates from the critical exponent $\beta\simeq0.4$ very rapidly.

In the right plot of \Fig{fig:phase1}, we present contour lines of the slope of the order parameter at finite pion mass in the finite density region. We show contour lines corresponding to deviations of the slope from the critical exponent by $1\%$, $2\%$, and $3\%$. Similar to the behavior observed for the slope of the correlation length, the contour lines exhibit a peak in the region where the chemical potential lies between 400 MeV and 500 MeV. This peak originates from the fact that the slope deviates upward from the critical exponent at low chemical potential and downward at high chemical potential, while the intermediate transition region enlarges the range over which the slope agrees with the critical exponent $\delta\simeq5$.

%
%

%
\begin{figure*}[t]
\includegraphics[width=0.48\textwidth]{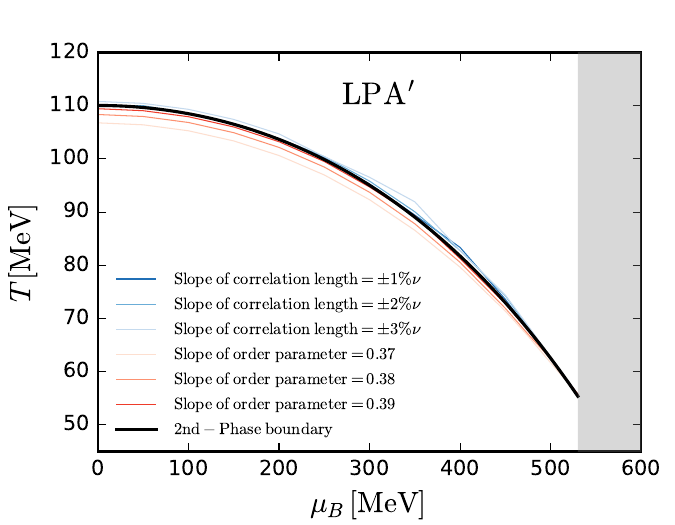}
\includegraphics[width=0.48\textwidth]{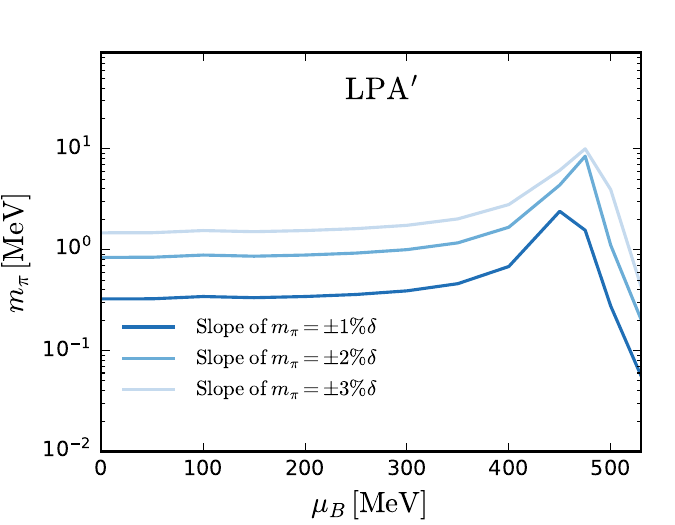}
\caption{Contour lines of the fitted $\mathrm{LPA}^\prime$ data slope on the phase diagram. In the left panel, the contour lines above the phase boundary (black solid line) indicate where the slope of the correlation length as a function of reduced temperature deviates from the critical exponent $\nu$ by $1\%$, $2\%$, and $3\%$. The lines below the phase boundary corresponding to the values 0.37, 0.38 and 0.39 of the slope of the order parameter $\sigma_0$ as a function of reduced temperature. In the right panel, the lines corresponding to $1\%$, $2\%$, and $3\%$ deviations of the slope of the order parameter as a function of pion mass from the critical exponent $\delta$. The gray area denotes regions where no data are available.}
\label{fig:phase2}
\end{figure*}
%

%
\begin{figure*}[t]
\includegraphics[width=0.48\textwidth]{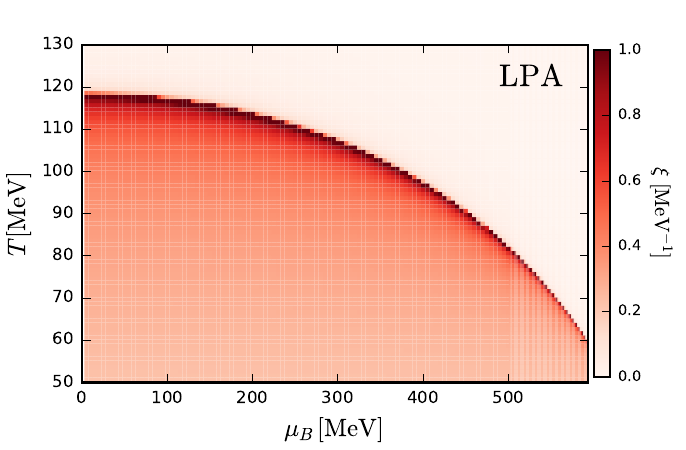}
\includegraphics[width=0.48\textwidth]{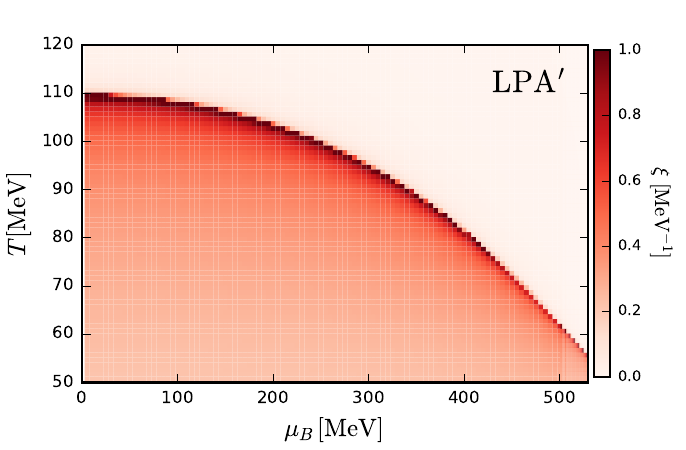}
\caption{Heatmap of the correlation length $\xi$ on the phase diagram of the QM model within LPA (left) and $\mathrm{LPA}^\prime$ (right).}
\label{fig:phase3}
\end{figure*}
%

In \Fig{fig:phase2}, we present the contour lines of the slopes obtained from the $\mathrm{LPA}^\prime$ calculation. The overall structure of the two figures is consistent with the LPA results, with only minor quantitative differences. For the slope of the correlation length in the temperature direction and the slope of the order parameter in the pion-mass direction, we continue to use contour lines corresponding to deviations of $1\%–3\%$ from the critical exponents $\nu\simeq0.759$ and $\delta\simeq4.79$. For the slope of the order parameter in the temperature direction, we choose the values 0.37, 0.38, and 0.39, a little smaller than $\beta\simeq0.397$. By comparing the results obtained in LPA and $\mathrm{LPA}^\prime$, we find that the contours in $\mathrm{LPA}^\prime$ lie closer to the phase boundary. The contours of $\sigma_0$ at finite pion mass also move to small values. These results indicate that after considering the meson anomalous dimension, the slope deviates more rapidly from the critical exponents.

To better visualize the behavior of the correlation length near the phase boundary, we show the heatmap plot of the correlation length on the phase diagram in \Fig{fig:phase3}. One can see that there is no qualitative difference between the LPA (left panel) and $\mathrm{LPA}^\prime$ (right panel) results. For a 2nd-order phase transition, the correlation length diverges at the critical temperature. In \Fig{fig:phase3}, the color scale of the heatmap is restricted to the range $0-1$ $\mathrm{MeV}^{-1}$ in order to more clearly display how the divergence of the correlation length evolves along the phase boundary. In the low temperature region, the correlation length remains finite. As the temperature increases toward the transition temperature, it tends to diverge, while in the high temperature region it decreases and approaches zero. The dark-colored region in the figure also indicates that the area where the correlation length divergence decreases as the chemical potential increases, consistent with the conclusion drawn earlier from the contour analysis.

From the analysis presented in this section, based on LO scaling fits to three different critical exponents, we find that with increasing chemical potential the QM model departs more rapidly from the regime that can be described by the $O(4)$ critical exponents of the chiral phase transition along both mass and temperature directions.
\section{Subleading scaling contribution}
\label{sec:subleading}
Up to this point, our discussion has been based on fits to the data that include only the leading contribution of the scaling functions, e.g., \Eq{eq:mpi_log,eq:t_log,eq:xi_log}. Therefore, we cannot exclude the possibility that subleading scaling corrections may modify our conclusions. In this section, we examine fits that incorporate next-to-leading-order (NLO) contributions to the scaling functions in order to verify that these subleading terms do not alter the range over which the slope remains consistent with the expected critical exponents.

Previous investigations of critical exponents in QCD \cite{Braun:2023qak} have provided the explicit form of the NLO contribution to the scaling function along the pion mass direction. The scaling form of the chiral condensate can be obtained from the rescaling of the free energy. The corresponding derivation can be found in \cite{Braun:2023qak,Zinn-Justin:2002ecy,Pelissetto:2000ek} and \Cref{app:sub_scaling}. Here we directly give the scaling form with NLO correction. Along the pion mass and temperature directions, we have
%
\begin{align}
\sigma_0(t=0,m^2_\pi)&=B_c\,m^{2/\delta}_\pi\,\big(1+a_m\,m^{2\theta_H}_\pi+\cdots\big)\,,\\[2ex]
\sigma_0(t,m^2_\pi=0)&=A\,|t|^{\beta}\,\big(1+a_t\,|t|^{\theta_t}+\cdots\big)\,,\\[2ex]
\xi(t,m^2_\pi=0)&=C\,|t|^{-\nu}\,\big(1+a_{\xi_t}\,|t|^{\theta_t}+\cdots\big)\,.
\end{align}
%
The NLO exponents can be obtained from the rescaling. The exponents are given by the combination of other exponents, e.g., $\theta_H=\nu\omega/(\beta\delta)$, $\theta_t=\nu\omega$. The exponent $\omega$ is the smallest irrelevant eigenvalue of the stability matrix. Note that, the NLO exponent is the same $\theta_t$ for the chiral condensate and correlation length. Due to the nonlinear form of the above equations, applying the linear slope-fitting procedure introduced previously does not yield sufficiently stable results. Therefore, we fix the NLO exponents $\theta_{H/t}$ and perform fits to the LO and NLO critical amplitudes ($B_c$, $A$, $C$, $a_m$, $a_t$ and $a_{\xi_t}$) as well as the LO critical exponents.

%
\begin{figure*}[t]
\includegraphics[width=0.3405\textwidth]{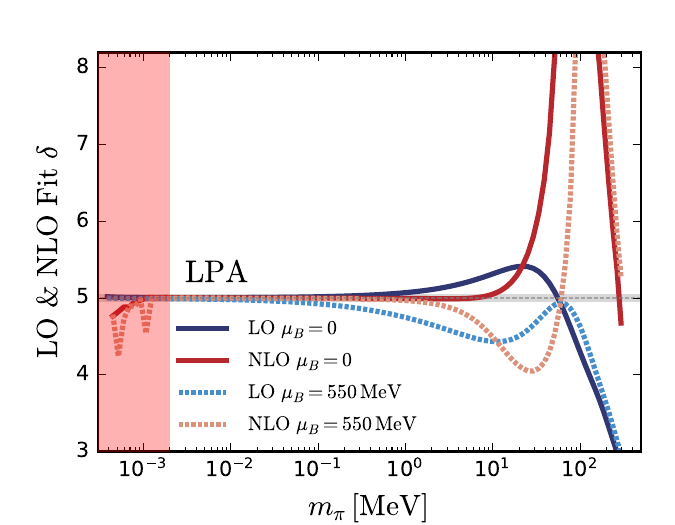}\hspace{-3mm}
\includegraphics[width=0.3405\textwidth]{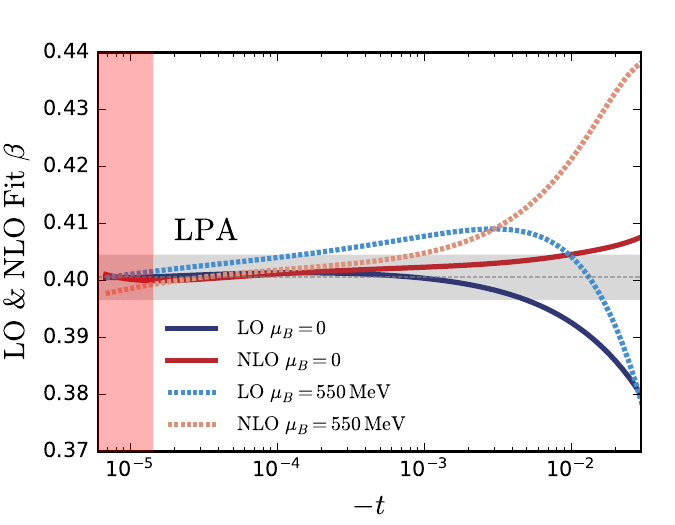}\hspace{-3mm}
\includegraphics[width=0.3405\textwidth]{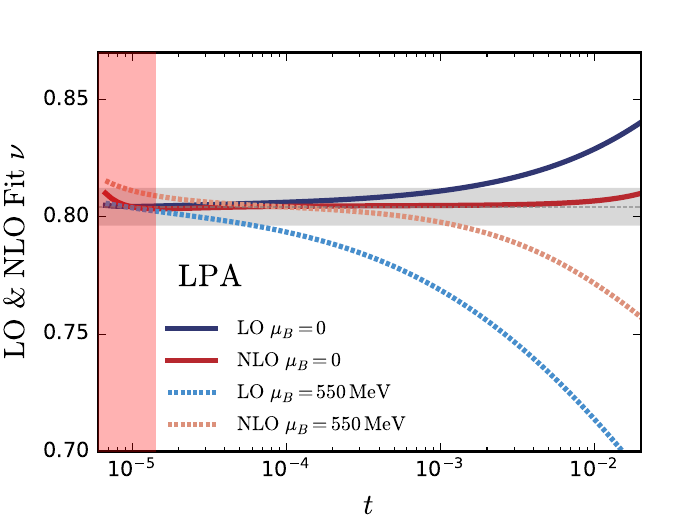}
\caption{Comparison of LO and NLO fits to the LPA data at $\mu_B=0$ and $\mu_B=550$ MeV. Left: leading exponent of the order parameter at finite pion mass.
Middle: leading exponent of the order parameter at finite reduced temperature.
Right: leading exponent of the correlation length at finite reduced temperature.}
\label{fig:NLO_LPA}
\end{figure*}
%

\begin{figure*}[t]
\includegraphics[width=0.3405\textwidth]{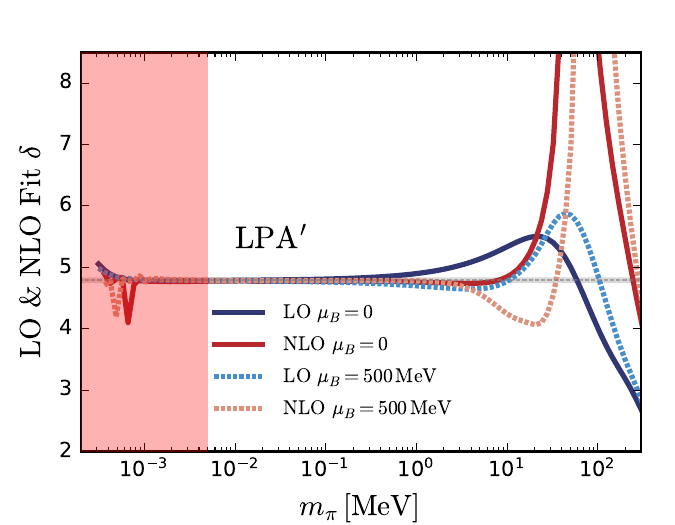}\hspace{-3mm}
\includegraphics[width=0.3405\textwidth]{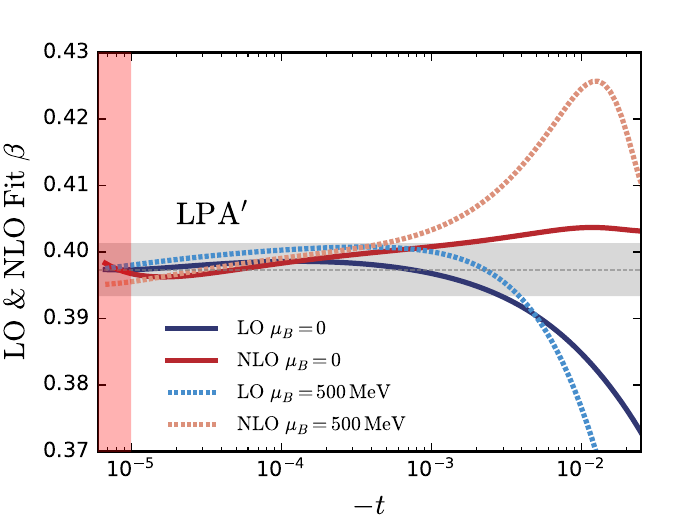}\hspace{-3mm}
\includegraphics[width=0.3405\textwidth]{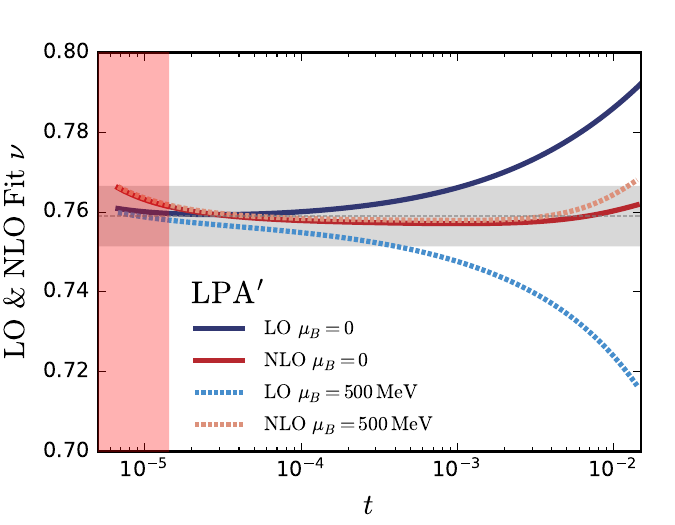}
\caption{Comparison of LO and NLO fits to the $\mathrm{LPA}^\prime$ data at $\mu_B=0$ and $\mu_B=500$ MeV. Left: leading exponent of the order parameter at finite pion mass.
Middle: leading exponent of the order parameter at finite reduced temperature.
Right: leading exponent of the correlation length at finite reduced temperature.}
\label{fig:NLO_LPAp}
\end{figure*}

In \Fig{fig:NLO_LPA} and \Fig{fig:NLO_LPAp}, we show the comparison of the LO and NLO fits along pion mass and temperature directions under LPA and $\mathrm{LPA}^\prime$ data. From these figures, one observes that at vanishing chemical potential, the NLO fits (blue solid lines) slightly extend the range over which the extracted exponents agree with the critical values compared to the LO fits (red solid lines) for all three critical exponents. In particular, for the exponent along the pion-mass direction, the NLO fit is consistent with the fRG-QCD result. The fitted exponent agrees with the critical value in the region $m_\pi\lesssim10$ MeV for both LPA and $\mathrm{LPA}^\prime$. It is also worth noting that the exponent associated with the temperature dependence of the order parameter (see middle plots of both figures) changes from a downward deviation to an upward deviation as one moves away from the critical temperature. This behavior originates from the subleading amplitude contribution, which modifies the exponent at larger reduced temperatures.

We further show the fitting results at larger chemical potentials in order to assess whether NLO contributions alter the tendency of the critical region to shrink at high chemical potential. For LPA we give the results at $\mu_B=550$ MeV, for $\mathrm{LPA}^\prime$ we give the results at $\mu_B=500$ MeV. After including the NLO contributions, the behavior of the exponents at high chemical potential remains consistent with that obtained from the LO analysis. The interval over which they agree with the critical values becomes smaller than at zero chemical potential. Therefore, the conclusions drawn in the previous section remain unchanged after NLO contributions are taken into account. Interestingly, at high chemical potential, the NLO fit to the temperature dependence of the correlation length (light brown dashed line) deviates downward in LPA, while in $\mathrm{LPA}^\prime$ it deviates upward. Since the only difference between LPA and $\mathrm{LPA}^\prime$ is the inclusion of the mesonic anomalous dimension, we infer that this change in the deviation pattern originates from the NLO exponent modified by the anomalous dimension.
\section{Summary and Conclusions}
\label{sec:conclusion}
%
%
In this work, we calculate the critical exponents of the QM model within the LPA and $\mathrm{LPA}^\prime$, and our results are consistent with those obtained in previous studies \cite{Chen:2021iuo,Litim:2001hk,Bohr:2000gp}. We extract the critical exponents $\delta$, $\beta$ and $\nu$ by fitting the model data to the scaling relation given in \Eq{eq:mpi_log,eq:t_log,eq:xi_log}. From these three equations, one can see that in the vicinity of the critical temperature in the chiral limit, the slope of the logarithmic data corresponds to the critical exponents. As the temperature and the pion mass move away from the second-order phase transition, the slope gradually deviates from the critical exponents, and eventually the critical scaling relation ceases to be valid beyond a certain temperature or mass range. We estimate how this range evolves with the chemical potential by fitting the numerical data. Finally, we examine how this behavior is modified once the NLO scaling corrections are included in the fits.

First, for the behavior of the chiral order parameter as a function of the pion mass, we fit the slope of the numerical results using the scaling relation. The slope approaches the critical exponent $\delta$ at the second-order transition temperature in the chiral limit and gradually deviates from this value as the pion mass increases. The pion mass range over which the slope deviates from $\delta$ is consistent with fRG calculations in QCD \cite{Braun:2023qak}. As the chemical potential increases, the slopes obtained in both the LPA and the $\mathrm{LPA}^\prime$ gradually change from deviating toward larger values than the critical exponent to deviating toward smaller values. In an intermediate range of chemical potential, the flat range of the slope becomes slightly enlarged. The deviation of the slope from the critical exponent in the $\mathrm{LPA}^\prime$ is slightly smaller than that in the LPA.

Next, for the behavior of the chiral order parameter as a function of temperature, we apply the same method to fit the slope of the data. We find that the slopes obtained in both the LPA and the $\mathrm{LPA}^\prime$ monotonically deviate toward smaller values from the critical exponent $\beta$. Moreover, as the chemical potential increases, the slope starts to deviate from $\beta$ at temperatures closer to the critical temperature. In other words, at high chemical potential the scaling behavior of the order parameter along the temperature direction is restricted to a small temperature interval around the critical temperature.

In the region above the phase transition temperature, we compute the slope of the logarithm of the correlation length with respect to the logarithm of the reduced temperature. As the temperature increases, the slope gradually deviates from the critical exponent $\nu$. Similar to the slope of the order parameter along the mass direction, the slope of the correlation length along the temperature direction changes, with increasing chemical potential, from deviating toward values larger than the critical exponent to deviating toward values smaller than the exponent. This also leads to a slight enlargement of the flat range for chemical potentials between 300 MeV and 500 MeV for both LPA and $\mathrm{LPA}^\prime$ results.

Finally, we perform fits including the NLO contributions to the scaling functions. We find that the inclusion of NLO correction can slightly enlarge the scaling range along both mass and temperature directions. At high chemical potential, the behavior of the flat range remains consistent with the LO analysis. It also shrinks as the chemical potential increases.

We now summarize the main results of the calculations presented above. In the QM model, when moving away from the second-order phase transition along the mass and temperature directions, the behavior of the order parameter and the correlation length can still be described by the scaling part of the thermodynamic potential within a relatively small range of mass and temperature. Beyond this range, the scaling function can no longer describe the behavior of the data. The region that can be described by the scaling function shrinks as the chemical potential increases. Similar to the results in \cite{Fu:2021oaw} that the crossover becomes increasingly sharp with increasing chemical potential, we find that the second-order chiral phase transition in the chiral limit also becomes sharper as the chemical potential increases. This also implies that if the applicability range of the scaling function is used to characterize the size of the critical region, the critical region of the model decreases with increasing chemical potential. Our results also show that for different critical exponents lead to different validity ranges of the scaling functions at finite chemical potential. Consequently, different observables yield different estimates for the size of the critical region. This suggests that, if one aims to quantify the static critical region by examining deviations of numerical fits from the expected critical exponents, it is necessary to take into account multiple observables rather than relying on a single one. Since the QM model does not include the full degrees of freedom of QCD, it cannot yet provide an estimate of the critical region in the real QCD system. However, it can be concluded that in a system with only the chiral transition of two light quark flavors, the critical region of the second-order phase transition shrinks rapidly at high chemical potential.

It should be noted that this work analyses the possible variation of the critical region at finite chemical potential solely from the perspective of critical exponents and does not rule out the possibility of differences arising from other methods of analysis. In the current computations of the meson effective potential, we use the Taylor expansion method. There are numerical challenges for calculations involving the TCP. Since the TCP may correspond to an unstable Gaussian fixed point, its critical region could be very small or even approach zero. We leave the investigation of the TCP to future work.
\section{Data availability}
The data are not publicly available. The data are available from the authors upon reasonable request.
\section*{Acknowledgements}

The author is grateful to Yong-rui Chen, Wei-jie Fu, Ugo Mire, Jan M. Pawlowski, Johannes V. Roth, Fabian Rennecke and Yang-yang Tan for valuable discussions.
This work is supported by the Alexander von Humboldt foundation.

\appendix 
\section{flow equations}
\label{app:flow_eq}

In the computation of the flow equations, we use the 3d-flat Litim regulator functions \cite{Litim:2000ci} for both boson and fermion
%
\begin{align}
R^\phi_k(\mathbf{q})&=Z_{\phi,k}\,\mathbf{q}^2\,r_\phi(\mathbf{q}^2/k^2)\,,\\[2ex]
R^q_k(\mathbf{q})&=Z_{q,k}\,i\,\bm{\slashed{q}}\,r_f(\mathbf{q}^2/k^2)\,.
\end{align}
%
The shape function is defined by the Heaviside theta function
%
\begin{align}
r_\phi(x)&=\bigg(\frac{1}{x}-1\bigg)\Theta(1-x)\,,\\[2ex]
r_f(x)&=\bigg(\frac{1}{\sqrt x}-1\bigg)\Theta(1-x)\,.
\end{align}
%
Then the dimensionless propagators of boson and fermion are as follows
%
\begin{align}
G_\phi(q,\bar m^2_\phi)&=\frac{1}{\tilde q_0^2+1+\bar m^2_{\phi,k}}\,,\\[2ex]
G_q(q,\bar m^2_q;\tilde \mu)&=\frac{1}{(\tilde q_0+i\tilde\mu)^2+1+\bar m^2_{q,k}}\,.
\end{align}
%
The $\tilde\mu=\mu/k$ and $\tilde q_0=q_0/k$ are the dimensionless quark chemical potential and Matsubara frequency.

The threshold functions which are used in the flow equation of effective potential are
%
\begin{align}
l_0^{(\phi,d)}(\bar m^2_{\phi,k},\eta_{\phi,k};T)&=\frac{2}{d-1}\bigg(1-\frac{\eta_{\phi,k}}{d+1}\bigg)\mathcal{B}_{(1)}(\bar m^2_{\phi,k};T)\,,\\[2ex]
l_0^{(q,d)}(\bar m^2_{q,k},\eta_{q,k};T,\mu)&=\frac{2}{d-1}\bigg(1-\frac{\eta_{q,k}}{d}\bigg)\mathcal{F}_{(1)}(\bar m^2_{q,k};T,\mu)\,.
\end{align}
%
In the computation we use $d=4$ and $\eta_{q,k}=0$. The definition of the boson and fermion threshold functions $\mathcal{B}_{(n)}$ and $\mathcal{F}_{(n)}$ are the Matsubara summation of the propagators
%
\begin{align}
\mathcal{B}_{n}(\bar m^2_{\phi,k},T)&=\frac{T}{k}\sum_{n_0}\Big(G_\phi(q,\bar m^2_{\phi,k})\Big)^n\,,\\[2ex]
\mathcal{F}_{n}(\bar m^2_{q,k},T,\mu)&=\frac{T}{k}\sum_{n_0}\Big(G_q(q,\bar m^2_{q,k})\Big)^n\,.
\end{align}
%
The lowest order of the functions are given by
%
\begin{align}
\mathcal{B}_{(1)}(\bar m^2_{\phi,k},T)&=\frac{1}{\sqrt{1+\bar m^2_{\phi,k}}}\bigg(\frac{1}{2}+n_b(\bar m^2_{\phi,k},T)\bigg)\,,\\[2ex]
\mathcal{F}_{(1)}(\bar m^2_{q,k};T,\mu)&=\frac{1}{2\sqrt{1+\bar m^2_{q,k}}}\\
\times\bigg(1-&n_q(\bar m^2_{q,k};T,\mu)-n_q(\bar m^2_{q,k};T,-\mu)\bigg)\,.
\end{align}
%
The definitions of the boson and fermion distribution functions are
%
\begin{align}
n_b(\bar m^2_{\phi,k};T)&=\frac{1}{\mathrm{exp}\Big[\frac{k\sqrt{1+\bar m^2_{\phi,k}}}{T}\Big]-1}\,,\\[2ex]
n_q(\bar m^2_{q,k};T,\pm\mu)&=\frac{1}{\mathrm{exp}\Big[\frac{k\sqrt{1+\bar m^2_{q,k}}\mp\mu}{T}\Big]+1}\,.
\end{align}
%

The flow equation of the meson anomalous dimension can be obtained by \Eq{eq:eta_phi} as follows
%
\begin{align}
\eta_{\phi,k}&=\frac{1}{6\pi^2}\bigg\{\frac{4}{k^2}\bar{\kappa}_k\,\big(\bar{U}_k''(\bar{\kappa}_k)\big)^2\mathcal{BB}_{(2,2)}(\bar m^2_{\pi,k},\bar m^2_{\sigma,k};T)\nonumber\\[2ex]
&+N_c\,\bar{h}_k^2\Big[-3\,\mathcal{F}_{(2)}(\bar m^2_{q,k};T,\mu)+8\,\mathcal{F}_{(3)}(\bar m^2_{q,k};T,\mu)\Big]
\bigg\}\,.
\end{align}
%
The double meson threshold function is given by
%
\begin{align}
&\mathcal{BB}_{(n_1,n_2)}(m_{b_1}^2,m_{b_2}^2;T)\nonumber\\[2ex]
=&\frac{T}{k}\sum_{n_0}\bigg(G_b(q,m^2_{b_1,k})\bigg)^{n_1}\bigg(G_b(q,m^2_{b_2,k})\bigg)^{n_2}\,.
\end{align}
%
The lowest order of the function is
%
\begin{align}
&\mathcal{BB}_{(1,1)}(m^2_{b_1},m^2_{b_2};T)\nonumber\\[2ex]
=&-\bigg\{\bigg(\frac{1}{2}+n_b(m^2_{b_1,k};T)\bigg)\,\frac{1}{\sqrt{1+m^2_{b_1,k}}(m^2_{b_1}-m^2_{b_2})}\nonumber\\[2ex]
&\,\,\,\,\,\,\,+\bigg(\frac{1}{2}+n_b(m^2_{b_2,k};T)\bigg)\,\frac{1}{\sqrt{1+m^2_{b_2,k}}(m^2_{b_2}-m^2_{b_1})}
\bigg\}\,.
\end{align}
%
The $m_{b_1}$ and $m_{b_2}$ give us different meson masses. The higher order of the threshold functions can be obtained by the derivative of the first order of the function with respect to the masses. For example
%
\begin{align}
&\mathcal{BB}_{(n_1+1,n_2)}(m^2_{b_1,k},m^2_{b_2,k};T)\nonumber\\[2ex]
=&-\frac{1}{n_1}\frac{\partial}{\partial\,m^2_{b_1,k}}\mathcal{BB}_{(n_1,n_2)}(m^2_{b_1,k},m^2_{b_2,k};T)\,.
\end{align}
%
\section{Numerical setup}
\label{app:numerical}
Here we introduce the numerical setup of the fRG flow equations. First, we set the UV cutoff scale of the theory to $\Lambda=500$ MeV. Because our model is a low energy effective theory, so the choice of a small cutoff scale is reasonable. This small scale is also used in, e.g., \cite{Chen:2021iuo}. The behavior of the critical exponents at finite chemical potential does not depend on the value of the cutoff scale.

For the effective potential, we have the form at UV scale as follow
%
\begin{align}
U_\Lambda(\rho)=\frac{\lambda_\Lambda}{2}\rho^2+\nu_\Lambda\,\rho\,.
\end{align}
%
For both LPA and $\mathrm{LPA}^\prime$, we give the parameter setup in \Tab{tab:param-set}.
%
\begin{table}[h]
  \begin{center}
 \begin{tabular}{c|ccc}
    \hline\hline & & & \\[-2ex]   
    & \hspace{0.5cm} $\Lambda=500$ $[\mathrm{MeV}]$& LPA & \hspace{0.5cm} $\mathrm{LPA}^\prime$ \\[1ex]
    \hline & &  \\[-2ex]
    & $\lambda_{\Lambda}$   & 5.7  & 5.7 \\[1ex]
   parameters & $\nu_{\Lambda}$ $[\mathrm{GeV}^2]$  & 0.290 & 0.384   \\[1ex]
    & \hspace{0.5cm}$c_\sigma$ $[\times10^{-3}\mathrm{GeV}^3]$\hspace{0.5cm}   & 1.7 & 2.1  \\[1ex]
    & $\bar h_\Lambda$  & 6.50 & 7.81  \\[1ex]
   \hline\\
    & \hspace{0.5cm} $\bar m_\pi$ $[\mathrm{MeV}]$ \hspace{0.5cm} & 135.62 & 137.47 \\[1ex]
    vacuum& \hspace{0.5cm} $\bar m_\sigma$ $[\mathrm{MeV}]$ \hspace{0.5cm} & 408.36 & 387.95 \\[1ex]
    results& \hspace{0.5cm} $\bar m_q$ $[\mathrm{MeV}]$ \hspace{0.5cm} & 300.41 & 300.47 \\[1ex]
    & \hspace{0.5cm} $\bar \sigma_0$ $[\mathrm{MeV}]$ \hspace{0.5cm} & 92.43 & 92.47 \\[1ex]
    \hline\hline
  \end{tabular}
  \caption{Parameters setup of the QM model with both LPA and $\mathrm{LPA}^\prime$.}
  \label{tab:param-set}
  \end{center}\vspace{-0.5cm}
\end{table}
%
In the $\mathrm{LPA}^\prime$ approximation, the initial condition of the meson wave function renormalization is $Z_{\phi,\Lambda}=1$.
\section{Critical exponents}
\label{app:CE}
%
\begin{figure}[t]
\includegraphics[width=0.48\textwidth]{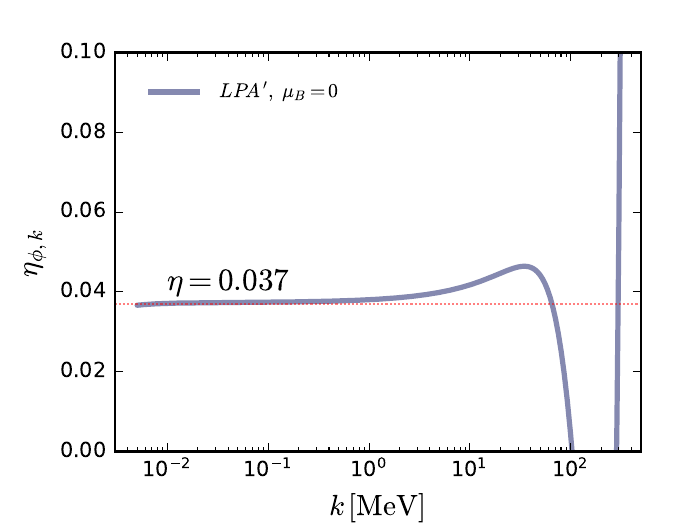}
\caption{The anomalous dimension as function of RG scale at vanishing chemical potential in $\mathrm{LPA}^\prime$ computation.}
\label{fig:eta}
\end{figure}
%
Previous studies have found that the numerical determination of critical exponents shows a mild dependence on the method used to solve the effective potential, see, e.g., \cite{Chen:2021iuo}. Here we also list our Taylor expansion results to make it convenient to compare with other methods.

In addition to the three critical exponents discussed in the main text, we also present the results for the anomalous dimension calculated within the $\mathrm{LPA}^\prime$ in \Fig{fig:eta}. Then we can give the table of exponents in \Tab{tab:exp}

%
\begin{table}[h]
  \begin{center}
 \begin{tabular}{ccccc}
    \hline\hline & & \\[-2ex]   
    \hspace{0.2cm} $\Lambda=500$ $[\mathrm{MeV}]$ & \hspace{0.5cm} $\beta$ & \hspace{0.5cm} $\delta$ & \hspace{0.5cm} $\nu$ & \hspace{0.5cm} $\eta$\\[1ex]
    \hline & &  \\[-2ex]
   LPA   & \hspace{0.5cm} 0.400 & \hspace{0.5cm} 5.00 & \hspace{0.5cm} 0.804 & \hspace{0.5cm} - \\[1ex]
   $\mathrm{LPA}^\prime$  & \hspace{0.5cm} 0.397 & \hspace{0.5cm} 4.79 & \hspace{0.5cm} 0.759 & \hspace{0.5cm} 0.037 \\[1ex]
    \hline\hline
  \end{tabular}
  \caption{Parameters setup of the QM model with both LPA and $\mathrm{LPA}^\prime$.}
  \label{tab:exp}
  \end{center}\vspace{-0.5cm}
\end{table}
%

In the NLO fits, we also make use of the next-to-leading critical exponent $\omega$. This exponent is obtained from the smallest eigenvalue of the stability matrix corresponding to the irrelevant parameter. For the details of the computation, see \cite{Braun:2023qak}. Here we directly give the values of them, $\omega_{\mathrm{LPA}}=0.7459$ and $\omega_{\mathrm{LPA}^\prime}=0.6745$.

\section{Subleading scaling corrections}
\label{app:sub_scaling}

The scaling forms for the temperature dependence of the order parameter and the correlation length can be derived by rescaling the free energy and the correlation function, respectively. The derivation of the scaling function for the pion-mass dependence of the order parameter follows that given in \cite{Braun:2023qak}.

We first consider the rescaling of the free energy 
%
\begin{align}
f(u_T,u_H,u_2,u_3)=s^{-d}f(u_T s^{\frac{1}{\nu}},u_H s^{\frac{\beta\delta}{\nu}},u_2 s^{-\omega},u_3 s^{-b_3})\,,
\end{align}
%
where the $\omega$ and $b_3$ are the exponents of the NLO and next-next-leading-order (NNLO) of the scaling function. If we choose $s=u_T^{-\nu}$ and take the derivative of the free energy with respect to the external field H, we can get the temperature scaling function of the order parameter
%
\begin{align}
\sigma_0=|t|^\beta\,f_T(z,\omega_2,\omega_3)\,.
\end{align}
%
Here we use $z=u_Hu_T^{-\beta\delta}$, $\omega_2=u_2 u_T^{\nu\omega}$ and $\omega_3=u_3 u_T^{\nu b_3}$ with $u_T\propto t$ and $u_H\propto H$. In the end, after an expansion around $t=0$, we have
%
\begin{align}
\sigma_0(t,H=0)=A|t|^\beta\Big(1+a_t|t|^{\nu\omega}+a_{t3}|t|^{\nu b_3}+\cdots\Big)\,.
\end{align}
%

For the correlation length, which already carries the dimension of length, one can perform the rescaling in the following form
%
\begin{align}
\xi=s^{-1}f_\xi(u_T s^{\frac{1}{\nu}},u_H s^{\frac{\beta\delta}{\nu}},u_2 s^{-\omega},u_3 s^{-b_3})\,.
\end{align}
%
We can choose the same rescaling factor as $s=u_T^{-\nu}$. Then we get
%
\begin{align}
\xi=|t|^{-\nu} f_\xi(z,\omega_2,\omega_3)\,.
\end{align}
%
After the expansion of the reduced temperature, we have
%
\begin{align}
\xi(t,H=0)=C\,|t|^{-\nu}\,\Big(1+a_{\xi_t}\,|t|^{\nu\omega}+a_{\xi_{t3}}|t|^{\nu b_3}+\cdots\Big)\,.
\end{align}
%

\vfill 

\bibliography{ref-lib}%

\end{document}